\documentclass[%
 reprint,
 amsmath,amssymb,
 aps,
]{revtex4-1}
\usepackage{graphicx}
\usepackage{dcolumn}
\usepackage{bm}
\usepackage{graphicx}
\usepackage{dcolumn}
\usepackage{bm}
\usepackage{bigints}
\usepackage{color}
\usepackage{amsmath,amssymb,graphicx}
\usepackage{float}
\usepackage{braket}
\usepackage{tabularx}
\usepackage{tabulary}
\usepackage{tabu}
\usepackage{booktabs}

\newcolumntype{C}{>{\centering\arraybackslash}X}

\def\omegab{\bar{\omega}}

\begin{document}
\preprint{APS/123-QED}
\title{Photon pair generation in multimode optical fibers via intermodal phase-matching}
\author{Hamed Pourbeyram$^{1,2}$}
\author{Arash Mafi$^{1,2}$}%
\email{mafi@unm.edu}
\affiliation{$^1$Department of Physics \& Astronomy, University of New Mexico, Albuquerque, NM 87131, USA \\
             $^2$Center for High Technology Materials, University of New Mexico, Albuquerque, NM 87106, USA} 

\begin{abstract}
We present a detailed study of photon-pair generation in a multimode optical fiber via nonlinear four-wave mixing and intermodal phase-matching. 
We show that in multimode optical fibers, it is possible to generate correlated photon pairs in different fiber modes with large spectral shifts from the pump 
wavelength, such that the photon pairs are immune to contamination from spontaneous Raman scattering and residual pump photons. We also show that it is possible
to generate factorable two-photon states exhibiting minimal spectral correlations between the photon pair components in conventional multimode fibers using
commonly available pump lasers. It is also possible to simultaneously generate multiple factorable states from different FWM processes in the same fiber 
and over a wide range of visible spectrum by varying the pump wavelength without affecting the factorability of the states. Therefore, photon-pair generation in 
multimode optical fibers exhibits considerable potential for producing state engineered photons for quantum communications and quantum information processing applications.
\end{abstract}
\maketitle
\section{Introduction}
Generation of photon pairs and preparing them in a specific quantum state, from highly entangled to factorable photon pairs, is the backbone of 
many quantum optical technologies~\cite{2007-Milburn,2006-Maccone-Giovannetti,2005-Walmsley-Raymer,1997-Zeilinger}. Heralded single photon sources where the detection of one photon heralds the imminent 
arrival of the second photon can be used for quantum information processing when the photon pair source is in a factorable state i.e. each photon 
is in a pure-state. Parametric down conversion (PDC)~\cite{1967-Byer} and four-wave mixing (FWM)~\cite{1975-Stolen} are the two main processes that are used
for photon pair generation. FWM uses $\chi^{(3)}$ Kerr nonlinearity to generate photon pairs inside an optical fiber with the advantage of
low-loss coupling to other fiber optic communication components~\cite{2002-Kumar-Fiorentino}.

Photon pair generation has been reported in single-mode~\cite{2008-Kumar-Altepeter}, dispersion-shifted~\cite{2006-Kumar-Lee-1,2009-Kanter-Wang,2005-Inoue-Takesue}, 
birefringent~\cite{2005-Kumar-Li-1,2009-Walmsley-Smith-1,2014-Lorenz-Fang-2,2016-Raymer-Smith}, and micro-structured optical 
fibers~\cite{2005-Russell-Rarity,2010-Silberhorn,2004-Kumar-Sharping,2007-Walmsley-GarayPalmett,2005-Migdall-Fan-1,2006-Alibart-Rarity}. 
There has been a recent surge of interest in nonlinear optics of multimode fibers because of their rich nonlinear and dispersive 
properties~\cite{2015-Wise-Wright,2015-Wise-Wright-PRL,2012-NL-Mafi,2013-Mafi-Pourbeyram}. Photon pair generation in multimode optical fibers 
based on the $\chi^{(3)}$ parametric process~\cite{1975-Stolen,1982-Stolen-Bjorkholm} offers many attractive properties. So far,
the opportunities to tailor the quantum state of photon pairs in the rich landscape of multimode optical fibers has not received the 
attention it deserves. Photon pair generation in a multimode setting has been reported earlier in a few cases, e.g. in birefringent 
fibers~\cite{2009-Walmsley-Smith-1,2014-URen-Cruz-Delgado}. However, quantum properties and purity analysis of photon-pair state
in multimode fibers has not been analyzed in the published literature.

In this manuscript, we will explore the generation of the photon pairs in a commercial grade multimode optical fiber (Corning~SMF-28 
in the visible spectrum). We develop and employ geometrical and graphical tools to study quantum correlations of the generated photon pairs, 
analyze the purity of the heralded single photons, and explore the possibility for generating pure-state heralded single photons. 
We would like to emphasize the key advantages of photon-pair generation in multimode fibers. First, the photon pairs 
can be generated with a large spectral shift from the pump so that they can evade the contamination from both the residual pump photons and 
the spontaneous Raman scattering (SRS) photons~\cite{1982-Stolen-Bjorkholm,2015-Mafi-Pourbeyram-OPEX,1975-Stolen}. Second, the generation of photons in different spatial 
modes can provide opportunities for spatial mode entanglement~\cite{2012-Kang}. Third, the presence of multiple modes allows simultaneous generation of
photon pairs at different wavelengths (phase-matching points) in the same fiber~\cite{2015-Mafi-Pourbeyram-OPEX}. Forth, here we have shown that the 
correlation of the photon pairs can be readily controlled from a fully factorable to an entangled state.
 
The theoretical results here suggest that it is possible to generate factorable photon pairs over a wide range of wavelengths in multimode fibers.
Pure heralded photons can be generated over a wide range of wavelengths by varying the pump wavelength without the need for dispersion engineering~\cite{2006-Russel,2001-Sharping-Kumar-2}.
The pair-generation process and state properties are more robust compared with those obtained from birefringent-based phase-matching, because the birefringence can be
easily affected by the external changes such as temperature variations~\cite{2000-Gisin,2006-Kumar-Liang}. The diversity of multiple phase-matchings using
different sets of modes makes it possible to generate highly-factorable to fully-entangled two-photon states in these fibers.
\section{Two-photon state}
In a spontaneous FWM process, two pump photons $E_{p}$ and $E_{p^\prime}$ are jointly annihilated to create a photon pair comprised of a signal photon 
$E_{s}$ and an idler photon $E_{i}$. In a multimode optical fiber each photon involved in the FWM process can belong to a different spatial mode 
of the optical fiber identified by subscripts $p$, $p^\prime$, $s$, and $i$. Generated signal and idler photons can be co- or cross-polarized with the pump.
Here, we only consider the co-polarized case because it has a nonlinear production efficiency of three time larger than the cross-polarized case~\cite{2007-Agrawal-Lin-1}. 
Following a standard perturbative approach the two-photon component of the quantum state generated by FWM process can be written as
\begin{equation}
\ket{\Psi}\propto\iint d\omega_s~d\omega_i~f(\omega_s,\omega_i)~\hat{a}^{\dagger}_s(\omega_s)~\hat{a}^{\dagger}_i(\omega_i)\ket{0},
\label{Eq:general-form-2}
\end{equation}
where $\hat{a}^{\dagger}_j(\omega)$ is the photon creation operator at wavelength $\omega$ and spatial mode $j$. 
The joint spectral amplitude (JSA) $f(\omega_s,\omega_i)$ is given by~\cite{2004-Kumar-Chen,2005-Kumar-Chen-1}
\begin{equation}
f(\omega_s,\omega_i)=\iiint dz~d\omega_{p}~d\omega_{p^\prime}~E_{p}(\omega_{p})~E_{p^\prime}(\omega_{p^\prime})~e^{i\kappa z}.
\label{Eq:JSA}
\end{equation}
$\kappa$ is the phase mismatch term that will be defined later in Eq.~\ref{Eq:kappa}. $E_p(\omega_p)$ is the spectral amplitude of the pump and is modeled as a Gaussian function, 
and is defined by the total pulse energy $\mathcal{E}_p$, pump carrier frequency $\omegab_p$ and pump bandwidth $\sigma_p$
\begin{equation}
E_p(\omega_p)=\frac{\sqrt{\mathcal{E}_p}}{\sigma_p\sqrt{\pi}}~\exp\left(-\frac{(\omega_p-\omegab_p)^2}{\sigma_p^2}\right).
\label{Eq:pump}
\end{equation}

In this manuscript, we assume a degenerate-frequency pump configuration where the two pump photons share the same spectral information. 
We also assume degenerate-spatial-mode pump configuration where the two pump photons travel in the same spatial mode of the optical fiber. 
The reason we consider only the {\em single-spatial-mode} pump configuration is that the intermodal phase-matching with a single-spatial-mode 
configuration results in a considerably larger spectral shift between the pump, signal and idler photons compared 
to the non-degenerate mode case and other phase-matching techniques~\cite{1981-Lin,1982-Stolen-Bjorkholm}. This would be essential in protecting the signal photons from the Stokes-shifted
SRS photons and would make it easier to filter out the residual pump photons.
Using the Gaussian pump spectral amplitude from Eq.~\ref{Eq:pump} and defining the spectral shifts from phase-matched frequencies of the 
signal ($\omegab_s$) and idler ($\omegab_i$) with
\begin{align}
\nu_s=\omega_s-\omegab_s, \quad \nu_i=\omega_i-\omegab_i, \quad \Omega=\omega_p-\omegab_p,
\label{Eq:shifts}
\end{align}
and using the conservation of photon energies
\begin{align}
2\Omega=\nu_s+\nu_i,
\end{align}
we can simplify the JSA from Eq.~\ref{Eq:JSA} as
\begin{equation}
f(\nu_s,\nu_i)\propto\int_0^L dz~\exp\left(-\frac{(\nu_s+\nu_i)^2}{2\sigma^2_p}\right)~\exp({i\kappa z}).
\label{Eq:F}
\end{equation}

We note that we have also implicitly assumed the photon energy conservation at the phase-matched wavelengths where 
$2\omegab_p=\omegab_s+\omegab_i$. 
For a regular optical fiber the transverse geometry is invariant in the longitudinal direction; therefore, the phase mismatch $\kappa$ (Eq.~\ref{Eq:kappa})
is independent of the $z$ parameter. In this situation, the $z$ integral can be carried out analytically and we obtain  
\begin{equation}
f(\nu_s,\nu_i)\propto~\phi(\nu_s,\nu_i)~\alpha(\nu_s,\nu_i),
\label{Eq:F-Final}
\end{equation}
where $\phi(\nu_s,\nu_s)$ and $\alpha(\nu_s,\nu_s)$ are the pump and phase-matching terms given by
\begin{subequations}
\begin{align}
\phi(\nu_s,\nu_i) &= {\rm sinc}\Big(\frac{\kappa L}{2}\Big)~\exp\Big(i\kappa L/2\Big),\\
\alpha(\nu_s,\nu_i) &=\exp\left(-\frac{(\nu_s+\nu_i)^2}{2\sigma^2_p}\right).
\end{align}
\label{Eq:F-elements}
\end{subequations}
A factorable state is defined to be a state that can be written as a product of two functions in the form 
of $S(\nu_s)I(\nu_i)$, where $S(\nu_s)$ and $I(\nu_i)$ depend only on the signal and idler frequencies, respectively.
\subsection{Phase-matching}
Proper phase-matching is a necessary condition for the formation of FWM inside an optical fiber~\cite{2013-Agrawal-Book}.
Phase-matching can be achieved by various methods, such as the near zero-dispersion wavelength (ZDW) phase-matching~\cite{2005-Kumar-Li-1,1992-Inoue,2011-Pinto-Ameida}, 
birefringence-based phase-matching~\cite{2014-Lorenz-Fang-2,2009-Walmsley-Smith-1}, and intermodal 
phase-matching~\cite{1982-Stolen-Bjorkholm,2015-Ramachandran-Demas,2016-Mafi-Nazemosadat,2015-Mafi-Pourbeyram-MDPI,2012-Xu-Cheng}. 

In this work, we focus on intermodal phase-matching using a single-spatial-mode pump configuration~\cite{1975-Stolen}, where both pump photons propagate in the same fiber mode.
We would like to emphasize again that the reason we have chosen the intermodal phase-matching with a single-spatial-mode configuration is that it results in a considerably 
larger spectral shift between the pump, signal and idler photons compared to the non-degenerate spatial mode case and other phase-matching techniques~\cite{1981-Lin,1982-Stolen-Bjorkholm}. 
Intermodal phase-matching relies on the difference between the dispersion properties of different modes in a multimode optical fiber, where unlike a single-mode fiber, 
phase-matching can be achieved without any assistance from the nonlinear phase shift. The fact that the phase-matched signal and idler wavelengths are set far from the
pump wavelength and are independent of the pump power allows one to vary the pump power and still operate comfortably in the spontaneous FWM regime. In contrast, decreasing the
pump power in a single-mode fiber that is phase-matched near the ZDW brings the signal and idler wavelengths uncomfortably close to the pump.

Spontaneous Raman scattering photons in a silica fiber are generated over an approximately 50~THz frequency bandwidth, red-shifted relative to the pump, with a peak at $\sim$13~THz.  
Using a suitable optical fiber with properly chosen modal dispersion, it is possible to generate sufficiently large spectral shifts
~\cite{1982-Stolen-Bjorkholm,2015-Mafi-Pourbeyram-OPEX,1981-Lin} in order to protect the FWM signal photons from the SRS pollution. We note that there have been other 
attempts aimed at reducing the effect of the SRS, e.g. by polarization filtering~\cite{2004-Kumar-Li}, mode-matched filtering~\cite{2011-Kumar-Huang}, or cooling down the 
fiber~\cite{2006-Kumar-Lee-1,2011-Peng-Zhou,2005-Inoue-Takesue}.

Assuming a single-spatial-mode pump configuration, the phase mismatch $\kappa$ can be written as
\begin{equation}
\kappa=\beta_p(\omega_p)+\beta_p(\omega^\prime_{p})-\beta_s(\omega_s)-\beta_i(\omega_i)-2\gamma P,
\label{Eq:kappa}
\end{equation}
where in $\beta_j(\omega_u)$, the index $j$ ($j\in s,i,p$) identifies the spatial modes of the pump,
signal and idler photons and $\gamma$ is the nonlinear parameter. The phase-matching condition can 
be written as
\begin{equation}
2\beta_p(\omegab_p)-\beta_s(\omegab_s)-\beta_i(\omegab_i)-2\gamma P=0,
\label{Eq:phase-matching}
\end{equation}
where $\omegab_u$ ($u\in s,i,p$) indicates the phase-matched carrier frequencies.
\begin{figure}[h]
\includegraphics[width=3.2 in]{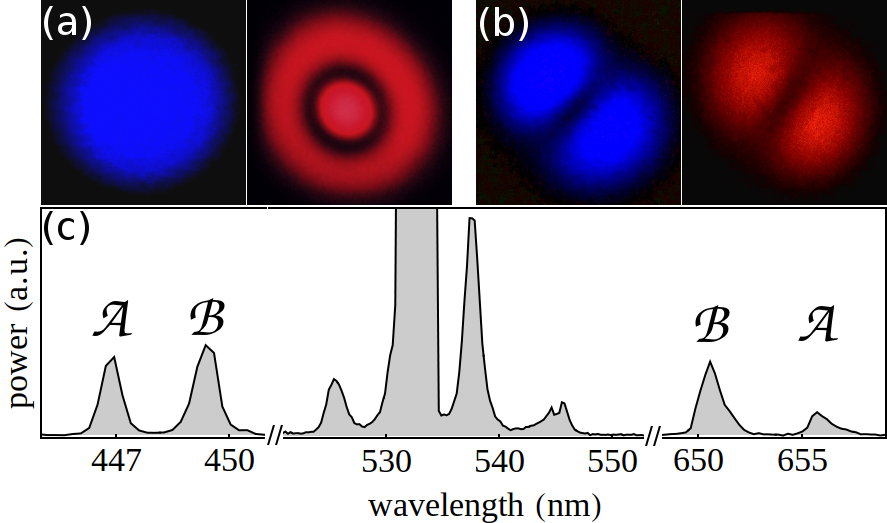}
\caption{Experimental observation of multiple FWM processes inside a short piece of SMF-28 fiber, showing the two most efficient processes generated far from the pump wavelength. 
(a) and (b) show the observed modes of FWM processes $\mathcal{A}$ and $\mathcal{B}$, respectively as will be discussed in Table.~\ref{table:eta}.}
\label{Fig:Experiment-smf28}
\end{figure}
\subsection{Inter-modal phase-matching in SMF-28}
\label{sec:intermodal1}
SMF-28 is a commercial grade optical fiber which is commonly used for single-mode optical fiber communications at 1550~nm wavelength. 
In the visible spectrum, this fiber supports a few spatial modes, so it is a multimode optical fiber. The theoretical calculations 
and numerical simulations we have done in this manuscript are in correspondence with the theoretical calculations and experimental 
observations we recently reported in Ref.~\cite{2015-Mafi-Pourbeyram-OPEX}. We observed the generation of multiple FWM processes 
in a 25~cm long piece of Corning SMF-28E+ fiber, which were triggered by both single-spatial-mode (degenerate) and non-degenerate-spatial-mode 
pump configurations using a high power 680~ps pulsed pump laser operating at 532~nm wavelength.

A pump in the fundamental ${\rm LP}_{01}$ mode of the fiber predominantly results in two FWM processes: The first process is observed 
with the signal and idler in ${\rm LP}_{02}$ and ${\rm LP}_{01}$ modes, respectively; and the second process is observed with the signal 
and idler both in ${\rm LP}_{11}$ modes. 
Here, ${\rm LP}_{mn}$ represents the spatial transverse field distribution of the guided linearly polarized mode in the step index 
fiber~\cite{Okamoto2006xiii}.
The first process which we will refer to as process $\mathcal{A}$ is usually the dominant output 
and has the farthest signal-idler peaks from the pump at 656~nm and 447~nm wavelengths. The second process which we will refer to as 
process $\mathcal{B}$ results in signal and idler peaks at 650~nm and 449~nm wavelengths. The measured spectrum and signal-idler spatial mode 
shapes for these two processes are shown in Fig.~\ref{Fig:Experiment-smf28}, where red (left) and blue (right) colors correspond to signal and idler modes
in each subfigure, respectively. 

\begin{figure}[h]
\includegraphics[width=3.3 in]{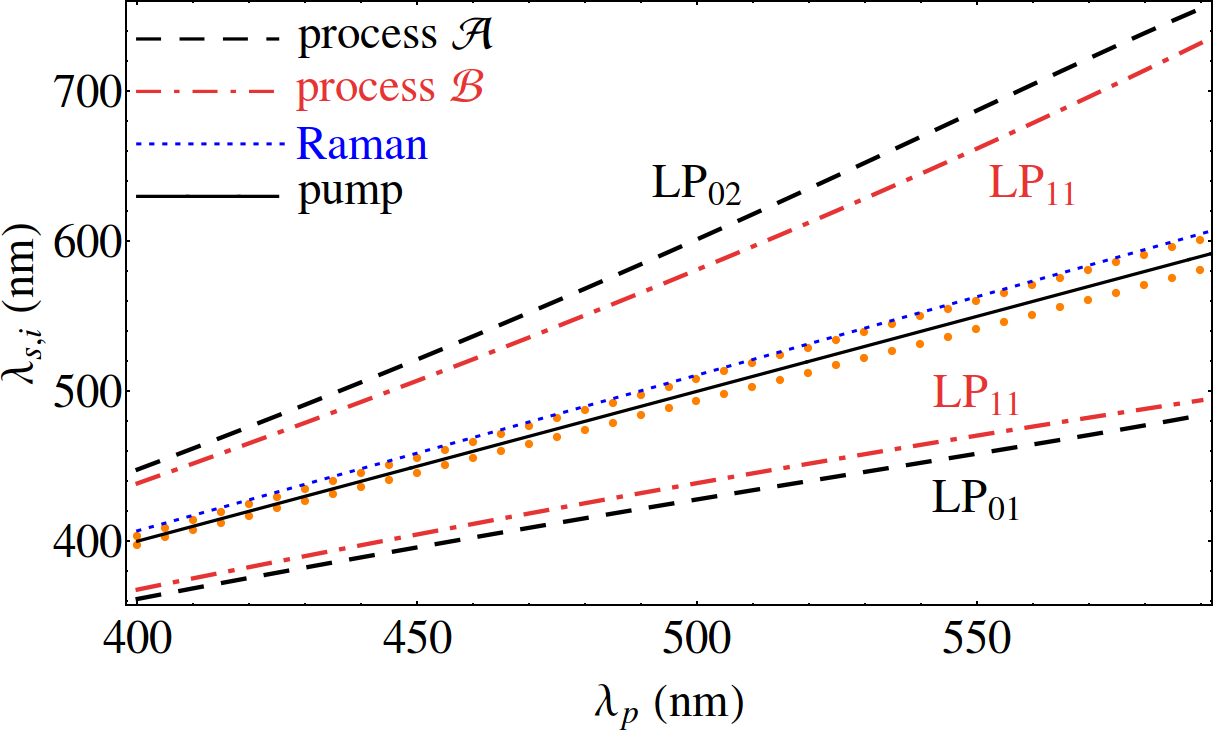}
\caption{Phase-matching curve as a function of the pump wavelength in SMF-28. The central diagonal solid black line represents the pump, and the
lines above and below belong to the signal and idler spectra. The two outer dashed black lines relate to process
$\mathcal{A}$ and the two dashed-dotted red lines relate to process $\mathcal{B}$. The orange dots adjacent to the pump represent a case of 
the non-degenerate-spatial-mode pump configuration, where the signal and idler are very close to the pump. The dotted blue line just above the pump 
indicates the SRS peak.}
\label{Fig:PM-smf28}
\end{figure}
In Fig.~\ref{Fig:PM-smf28}, we have plotted the phase-matching curves as a function of pump wavelengths in SMF-28 showing both degenerate- and  
non-degenerate-spatial-mode pump configurations. 
The case with the signal and idler pair close to the pump (orange dots) is for the non-degenerate-spatial-mode 
pump configuration with the pump photons propagating in ${\rm LP}_{01}$ and ${\rm LP}_{02}$ modes and the signal and idler photons propagating 
in ${\rm LP}_{02}$ and ${\rm LP}_{01}$ modes, respectively. We emphasize that the small spectral separations between the pump, signal, and idler is a characteristic
behavior of the FWM in the non-degenerate-spatial-mode pump configuration as discussed in detail in Ref.~\cite{1982-Stolen-Bjorkholm}. 

The two other FWM processes are far apart from pump (black-dashed and red-dashed-dotted) and are generated using a single-spatial-mode pump (${\rm LP}_{01}$) configuration. 
The dotted blue line shows the peak of the SRS and indicates the utility of the degenerate-pump-configuration to spectrally separate the signal-idler peaks from the SRS
contamination. 
\section{Quantum state purity}
\label{sec:purity}
The purity of a quantum system identified by the density matrix $\rho$ is defined as ${\rm Tr}(\rho^2)$, whose value is
between zero and one. If the quantum system is in a pure state $\ket{\Psi}$, then $\rho=\ket{\Psi}\bra{\Psi}$ and the 
purity equals one. For the two-photon signal-idler quantum state, the detection of one signal (idler) photon heralds 
the imminent arrival of the second idler (signal) photon and one would like to know whether the second idler (signal) 
photon is in a pure or a mixed state. This can be achieved by calculating the purity using the reduced density matrix
of the second idler (signal) state. The purity is given by $\mathcal{P}={\rm Tr}(\rho_i^2)={\rm Tr}(\rho_s^2)$, where 
$\rho_i={\rm Tr}_s(\rho)$ and $\rho_s={\rm Tr}_i(\rho)$ are the reduced density matrices of the idler and signal,
and are obtained by taking a partial trace over the signal and idler subsystems, respectively. We emphasize that the
purity value is the same for both signal and idler photons.

The two-photon state in Eq.~\ref{Eq:general-form-2} can be used to calculate the purity $\mathcal{P}$ as
\begin{equation}
\mathcal{P}=\dfrac{ \Big(\prod^4_{i=1} \int_{-\infty}^\infty d\nu_i\Big)
f_{\nu_1,\nu_2}
f^\ast_{\nu_3,\nu_2}f_{\nu_3,\nu_4}f^\ast_{\nu_1,\nu_4}}{\Big(\int_{-\infty}^\infty d\nu_1d\nu_2~|f_{\nu_1,\nu_2}|^2\Big)^2}.
\label{Eq:purity-4fIntegral}
\end{equation}
$f_{\nu_i,\nu_j}$ is the JSA defined in Eq.~\ref{Eq:F} whose frequency arguments are used as subscripts 
in Eq.~\ref{Eq:purity-4fIntegral} to reduce the size of the equation. The denominator in Eq.~\ref{Eq:purity-4fIntegral}
accounts for normalization of the JSA. In general, purity can be evaluated numerically using the matrix form
\begin{equation}
\mathcal{P}=\dfrac{{\rm Tr}(F.F^\dagger)^2}{{\rm Tr}^2(F.F^\dagger)},
\end{equation}
where $F$ is a matrix which represents the discrete version of the JSA function
$f({\nu_s,\nu_i})$ sampled uniformly over the frequency arguments $\nu_s$ and $\nu_i$;
therefore, the rows represent the signal and columns represent the idler frequencies.

It is possible to obtain a considerable analytical simplification of the expression for purity
by using Eqs.~\ref{Eq:F-Final} and~\ref{Eq:F-elements} and noting that the main contribution of the 
sinc function to the purity calculation in Eq.~\ref{Eq:purity-4fIntegral} comes from its central lobe, 
around which the sinc can be approximated as a Gaussian of the following form 
\begin{equation}
{\rm sinc}(x)\approx \exp(-\alpha x^2), \quad \alpha\approx 0.193.
\label{Eq:approx-sinc}
\end{equation}
We carry out a first-order Taylor expansion of the  
propagation constants in frequency around the phase-matched frequencies. Using the photon energy
conservation $\omegab_p+\omegab^\prime_{p}=\omegab_s+\omegab_i$, we obtain an approximate expression 
for $\kappa$ in the form
\begin{align}
\kappa L\approx \nu_s \tau_s+\nu_i \tau_i,
\label{Eq:PM}
\end{align}
where $\tau_s$ and $\tau_i$ are the group delays between the pump-signal and pump-idler photons defined by
\begin{subequations}
\begin{align}
\tau_s&=L\left[\beta^{(1)}_p(\omegab_p)-\beta^{(1)}_s(\omegab_s)\right],\\
\tau_i&=L\left[\beta^{(1)}_p(\omegab_p)-\beta^{(1)}_i(\omegab_i)\right].
\end{align}
\label{Eq:taus}
\end{subequations}
$\beta^{(1)}$ in Eq.~\ref{Eq:taus} signifies the first derivative of the propagation constant relative to the frequency argument.
It should be noted that Eq.~\ref{Eq:PM} is valid only for the single-spatial-mode pump configuration. From Eq.~\ref{Eq:F-Final} and 
Eq.~\ref{Eq:PM}, it is clear that $\tau_i$, $\tau_s$, the fiber length $L$, and the pump bandwidth $\sigma_p$ is all we 
need to have to construct the JSA.
By implementing Eq.~\ref{Eq:approx-sinc} and Eq.~\ref{Eq:PM} in the Eq.~\ref{Eq:F-elements}, JSA can be rewritten as
\begin{align}
\nonumber
f(\nu_s,\nu_i)&\propto\exp{\Big(i(\nu_s \tau_s+\nu_i \tau_i)L/2\Big)}\times\\
&\exp{\Big(-(\nu^2_s\tau^2_1+\nu^2_i\tau^2_2+2\,c\,\nu_s\nu_i\tau_1\tau_2)\Big)},
\label{Eq:F-approx}
\end{align}
given
\begin{subequations}
\begin{align}
\tau_1^2&=\frac{1}{2\sigma_p^2}+\frac{\alpha\tau_s^2}{4},
\label{Eq:tau1s}\\
\tau_2^2&=\frac{1}{2\sigma_p^2}+\frac{\alpha\tau_i^2}{4},
\label{Eq:tau2i}\\
c\,\tau_1\tau_2&=\frac{1}{2\sigma_p^2}+\frac{\alpha\tau_s\tau_i}{4}.
\label{Eq:c-t1t2-termc}
\end{align}
\label{Eq:c-t1t2}
\end{subequations}
From Eq.~\ref{Eq:F-approx} it is clear that if $c\,\tau_1\tau_2=0$ then JSA would be factorable in $\nu_s$ and $\nu_i$, which corresponds to the absence of correlations between the frequencies of the idler and signal photons. This could be achieved only if both terms are equal to zero or $\tau_s\tau_i<0$. The latter means that one of signal and idler photons should travel faster than pump photons while the other one travel with a group velocity smaller than pump photons~\cite{2007-Walmsley-GarayPalmett}.
After a few line of algebra we have shown that purity for the JSA given 
by Eq.~\ref{Eq:F-approx} would be 
\begin{equation}
\mathcal{P}=\sqrt{1-c^2}.
\label{Eq:purity-theory}
\end{equation}
This means that for JSA of Eq.~\ref{Eq:F-approx} one can find the purity by only knowing the parameters given by pump bandwidth, fiber length, 
and dispersion parameters of the optical fiber using Eq.~\ref{Eq:c-t1t2-termc}.
\subsection{Joint spectral amplitude ellipse}
The joint spectral amplitude has an elegant geometrical description based on the elliptical form 
of its magnitude in Eq.~\ref{Eq:F-approx}. The ellipse describes the contour of the maximum 
amplitude of JSA in the shifted frequency coordinates $\nu_s$ and $\nu_i$. Using this geometrical 
description, we will argue that the state purity is only a function of the shape of the ellipse 
and its orientation relative to the $\nu_s$-$\nu_i$ axes and the size of the ellipse does not affect 
the purity. 

\begin{figure}[htb]
\includegraphics[width=1.9in]{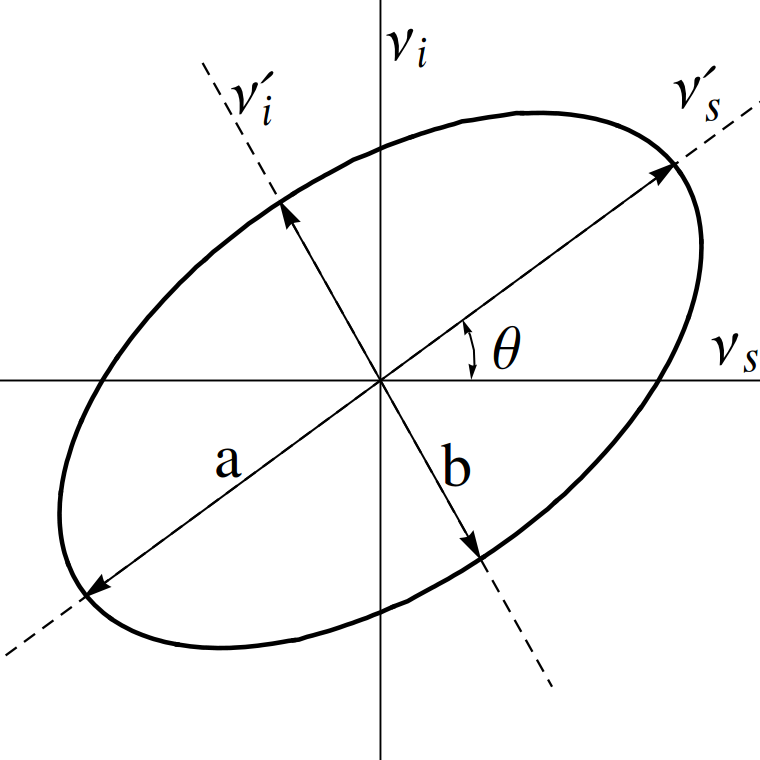}
\caption{Purity ellipse: a and b are the major and minor axes, respectively, and $\theta$ is the angle between two coordinates $(\nu_s,\nu_i)$ and $(\nu^{\prime}_s,\nu^{\prime}_i)$.}
\label{fig:purity-ellipse}
\end{figure}

Using a coordinate transformation (rotation by angle $\theta$) of the form
\begin{subequations}
\begin{align}
\nu^\prime_s&=\nu_s\cos\theta-\nu_i\sin\theta,\\
\nu^\prime_i&=\nu_s\sin\theta+\nu_i\cos\theta,
\end{align}
\end{subequations}
it is possible to convert the elliptical shape in Eq.~\ref{Eq:F-approx} to the following factorable form
\begin{equation}
f(\nu^\prime_s,\nu^\prime_i)=\mathcal{A}~\exp\left(-\nu^{\prime 2}_s/a^2\right)\exp\left(-\nu^{\prime 2}_i/b^2\right).
\label{Eq:function-pure}
\end{equation}
The JSA in new coordinates is identified by the angle $\theta$ and the ellipse semi-major and -minor axes $a$ and $b$, respectively. 
These parameters are related to the parameters in Eq.~\ref{Eq:c-t1t2} ($\tau_1$, $\tau_2$ and $c$) by
\begin{subequations}
\begin{align}
\tau_1^2&=\frac{\cos^2\theta}{a^2}+\frac{\sin^2\theta}{b^2},
\label{Eq:tau1ab}\\
\tau_2^2&=\frac{\cos^2\theta}{b^2}+\frac{\sin^2\theta}{a^2},
\label{Eq:tau2ab}\\
c\,\tau_1\tau_2&=\sin\theta\cos\theta\big(\frac{1}{b^2}-\frac{1}{a^2}\big).
\end{align}
\label{Eq:rotated-coordinates}
\end{subequations}
This coordinate transformation is shown in Fig.~\ref{fig:purity-ellipse}.

Solving Eq.~\ref{Eq:rotated-coordinates} for $c$ and plugging it in Eq.~\ref{Eq:purity-theory} 
gives us the purity as a function of the geometrical parameters of the JSA ellipse
\begin{equation}
\mathcal{P}=\frac{1}{\sqrt{1+(r-1/r)^2\sin^2\theta\cos^2\theta}}, \quad r=\dfrac{b}{a}.
\label{Eq:p}
\end{equation}
This result implies that purity is only determined by the ellipse shape parameter, i.e. its aspect ratio $r$
and the its orientation angle $\theta$. Therefore, the ellipse area $\pi a b$ can take any value and as long as
the aspect ratio and the orientation angle remain unchanged, the purity remains the same.

From Eq.~\ref{Eq:p}, it can be shown that in order to get a factorable two-photon state with $\mathcal{P}=1$,
it is sufficient to have $\theta=0$ or $\theta=\pi/2$ which means that the ellipse axes are aligned with the
shifted frequency coordinates $\nu_s$ and $\nu_i$. This is clear because in this case,
the JSA in Eq.~\ref{Eq:F-approx} becomes factorable to two independent Gaussians in $\nu_s$ and $\nu_i$ and 
$c=0$ as expected. We also note that from Eq.~\ref{Eq:p}, it is possible to obtain $\mathcal{P}=1$ when 
$r=1$. This is the limit where the ellipse turns into a circle and because the circle does not have 
a preferred orientation, one would naturally expect to obtain the same result for $\theta=0,\pi/2$. 

For the case of $r=1$, one can use Eqs.~\ref{Eq:tau1s},~\ref{Eq:tau2i},~\ref{Eq:tau1ab}, and~\ref{Eq:tau2ab} to show that
it results in $\tau^2_1=\tau^2_2$ and $\tau_s=-\tau_i$, where $c=0$. Therefore, the group velocity mismatch between 
pump-signal and pump-idler photons are equal and with opposite signs, resulting in a symmetric temporal separation 
of the signal and idler from the pump. In other words, the idler is delayed and signal is advanced relative to the pump 
with the same amount or vice versa. It is also possible to obtain the purity of equal to one in an extreme case 
when $r\to\infty$ or $r\to 0$. This situation requires either $\tau_i\to 0$ or $\tau_s\to 0$, in addition to 
having a ultra-broadband pump $\sigma_p\to\infty$.

From a practical point of view, the most interesting scenario for purity of equal to one with $c=0$ corresponds to the case
where $\tau_s\tau_i=-2/\alpha\sigma^2_p$. Because $\tau_s$ and $\tau_i$ are proportional to the fiber lengths, this means that
for a given FWM process in an optical fiber, it is possible to either tune the length of the fiber or the pump bandwidth
to obtain purity of one, resulting in a factorable two-photon state. We note that our discussions so far have relied on the
Gaussian approximation to the sinc function in Eq.~\ref{Eq:approx-sinc}. There are certain differences between the purity 
calculated from the sinc function and its Gaussian approximation that will be discussed in detail in the next section using the 
purity contours.
\subsection{Purity contours}
In the previous section, we argued that the purity of a two-photon quantum state is only a function of the dimensionless
parameters that identify the shape and the orientation of the purity plot. These dimensionless parameters are constructed 
from the physical attributes of the system including the pump bandwidth, fiber length, and dispersion parameters. We define 
two dimensionless parameters $r_1$ and $r_2$ of the form 
\begin{equation}
r_1=\frac{\tau_s}{\tau_i}, \quad r_2=\frac{\sigma_{si}}{\sigma_p},\quad \sigma_{si}=1/\sqrt{\tau_s^2+\tau_i^2},
\end{equation}
where $r_1$ is the ratio of group velocity mismatch between pump-signal and pump-idler photons and is independent of the fiber length 
or the pump bandwidth. Therefore, $r_1$ is exclusively determined by the design of the fiber refractive index profile which determines 
the phase-matching wavelengths and the dispersion parameters. On the other hand, $r_2$ is the ratio of $\sigma_{si}$ to the pump bandwidth;
therefore, it is inversely proportional to the fiber length. We will show that these two parameters are sufficient to determine the value of 
the purity.

The choice of $r_1$ and $r_2$ to parametrize the purity is very useful in practice. Because $r_1$ is exclusively a function of the refractive 
index profile of the fiber, it is only determined at the fiber design stage and cannot be manipulated during an experiment. Of course, this
argument is based on the assumption that the experimentalist has already chosen a fixed pump wavelength as is common in most practical scenarios, because
for different pump wavelengths $r_1$ takes different values. The parameter that can be manipulated readily in an experiment is $r_2$, either
by changing the fiber length or by modifying the pump bandwidth using a spectral filter. This separation becomes instrumental in designing an 
experiment to target a desired value of purity or set the expectations of what is or is not possible to achieve in an experiment.

It follows from Eq.~\ref{Eq:purity-theory} that the purity can be written in the following form 
\begin{equation}
\mathcal{P}=\sqrt{\frac{2\,\alpha\,r_2^2\,(X_1-2)X_1}{\Big(2\,r_2^2\,X_1+\alpha\, r_1\Big)\Big(2\,r_2^2\,X_1+\alpha/r_1\Big)}},
\label{eq:finalPurity}
\end{equation}
where we have defined 
\begin{equation}
X_1=r_1+1/r_1.
\end{equation}
It should be noted that Eq.~\ref{Eq:purity-theory} is based on the Gaussian approximation of the sinc function. It should also be mentioned that $X_1$ 
is invariant under $r_1\to 1/r_1$ transformation; therefore, it is sufficient to study the purity only in the range of $-1\le r_1 \le 1$. It is also obvious 
that $r_2\ge 0$.

\begin{figure}[h]
\centering
\includegraphics[width=3.2in]{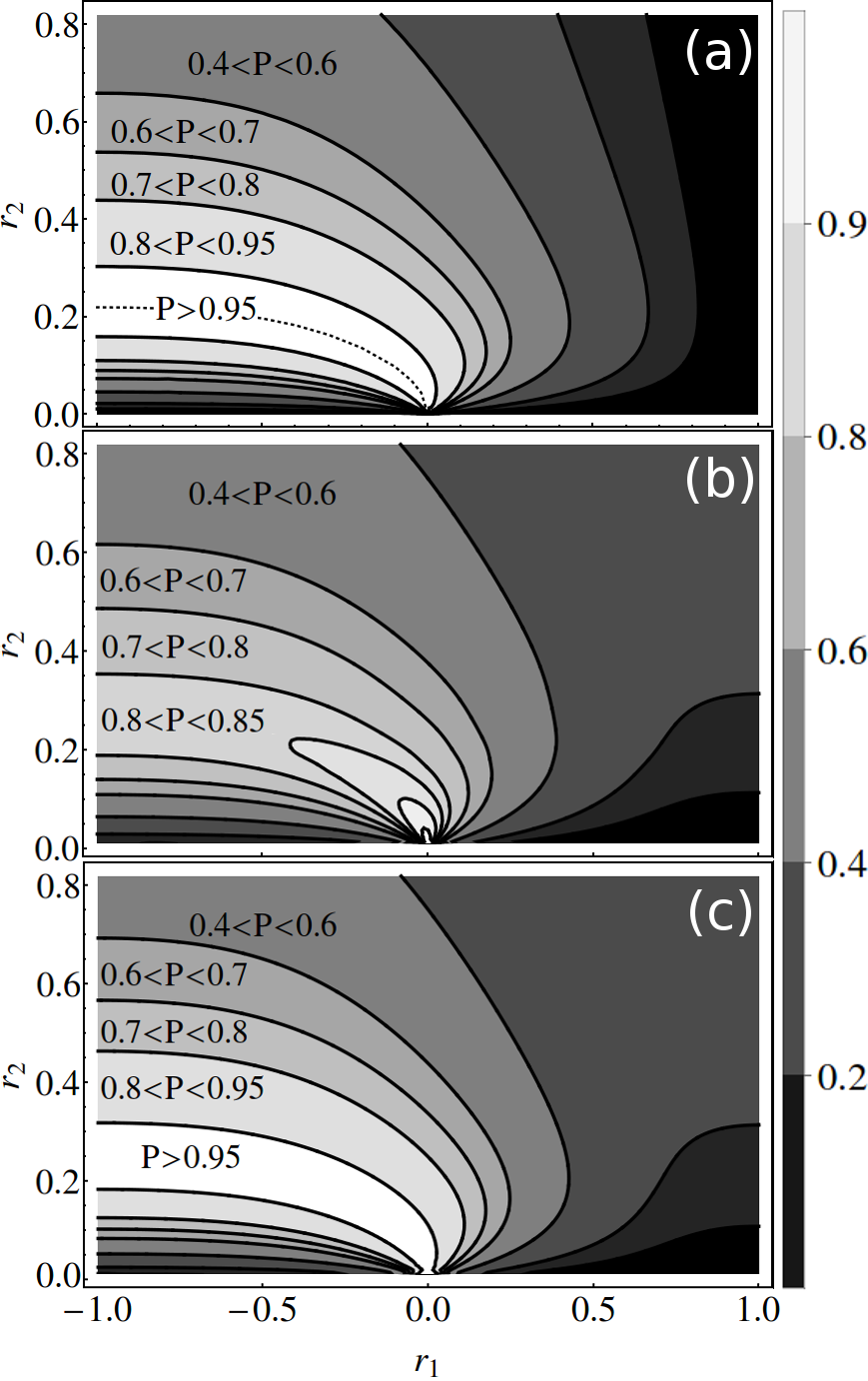}
\caption{Purity contours: (a) shows the purity contours calculated using Eq.~\ref{eq:finalPurity} with the Gaussian approximation. (b) and (c) show the purity contours evaluated
 using  Eq.~\ref{Eq:F-elements} and Eq.~\ref{Eq:purity-4fIntegral} without the Gaussian approximation, where in (c) a boxcar function ranging from $-\pi$ to $\pi$ has been used to filter out the side-lobes of the sinc function.}
\label{Fig:purity-r1-r2}
\end{figure}

In Fig.~\ref{Fig:purity-r1-r2}, we present contour plots of purity as a function of $r_1$ and $r_2$. Fig.~\ref{Fig:purity-r1-r2}(a) is determined by
Eq.~\ref{eq:finalPurity}, which is constructed from the JSA of Eq.~\ref{Eq:F-approx} with the Gaussian approximation to the sinc function. Using
Eq.~\ref{eq:finalPurity}, we determine that the contour corresponding to the purity value of one is parametrized as 
\begin{equation}
r_2^2(r_1+1/r_1)=-\alpha/2.
\label{eq:contourone}
\end{equation}
The contour line corresponding to the factorable two-photon state in Eq.~\ref{eq:contourone} is shown in Fig.~\ref{Fig:purity-r1-r2}(a) 
as a dotted line. We note that because $-\alpha/2< 0$, the contour representing 
the purity of one only lies in the $r_1<0$ region.

In Fig.~\ref{Fig:purity-r1-r2}(b), we show a contour plot of the purity which is calculated directly using Eq.~\ref{Eq:purity-4fIntegral} 
for the JSA of Eq.~\ref{Eq:F-Final} and Eq.~\ref{Eq:F-elements} without approximating the sinc function with a Gaussian form. While this 
contour plot closely resembles the one in Fig.~\ref{Fig:purity-r1-r2}(a), one can clearly see important differences, especially for purity 
values that are close to one. In particular, unlike the case with the Gaussian approximation, Fig.~\ref{Fig:purity-r1-r2}(b) clearly indicates 
that both $r_1$ and $r_2$ must be close to zero in order to obtain a factorable two-photon state. This visual comparison is essential because
it shows the impact of the Gaussian approximation that is widely used in the literature on the prediction of the purity of a heralded
single-photon state. We remind the reader that because of the implicit $r_1\to 1/r_1$ invariance of the purity, it is equally possible
to obtain a pure state with $r_2\to 0$ and $r_1\to \infty$. $r_2\to 0$ means that either pump needs to be very broadband or the fiber must be
sufficiently long~\cite{2009-Walmsley-Cohen}. On the other hand $r_1\to 0$ or $r_1\to \infty$ simply means that either the signal or idler must travel with the same group
velocity as the pump.

The main source of difference between the purity contours in Fig.~\ref{Fig:purity-r1-r2}(a) and Fig.~\ref{Fig:purity-r1-r2}(b) is the side-lobes in the sinc
function that are not accounted for in the Gaussian approximation to the sinc function in Eq.~\ref{Eq:approx-sinc}. The side-lobes of the sinc function 
are likely not interesting given the fact that they are spectrally farther from the phase-matched wavelengths. Once the idler and signal are separated 
by a diffraction grating, the side-lobes will not be easily coupled into the optical fibers that will carry the signal or idler into the photon 
counting detectors. Filtering out the side-lobes results only in 9.72\% reduction in the total flux. In Fig.~\ref{Fig:purity-r1-r2}(c), we show a contour 
plot of the purity, exactly similar to that of Fig.~\ref{Fig:purity-r1-r2}(b), except with the side-lobes of the sinc function removed by a bandpass 
box filter. The purity contour plot with the filtered sinc function strongly resembles the one with the Gaussian approximation in Fig.~\ref{Fig:purity-r1-r2}(a).
In particular, a large window opens up for purity near one around the same contour line of Eq.~\ref{eq:contourone}. Therefore, by sacrificing merely
9.72\% of the flux, it is possible to open up the opportunity to many realistic design scenarios to achieve a factorable two-photon state using the FWM process
in an optical fiber.
\section{Diversity of phase-matched FWM processes}
The presence of multiple spatial modes gives rise to the possibility to satisfy multiple phase-matched FWM processes
at the same time in a multimode fiber as previously shown in Fig.~\ref{Fig:PM-smf28}. These phase-matching conditions 
are each satisfied at different wavelengths and provide an opportunity to generate simultaneous photon pairs at different
wavelengths in the same fiber. However, not all processes are generated with the same efficiency. In practice, only a few 
of the possible phase-matched FWM processes are observed in an experiment. In the stimulated FWM scenarios, one or more
processes can totally dominate and suppress the others. However, for photon-pair generation which is a spontaneous 
process, the relative efficiency of a phase-matched FWM process is determined by a totally symmetric nonlinear coupling 
term defined as~\cite{2015-Mafi-Pourbeyram-OPEX}
\begin{equation}
\eta_{ppsi}=\Big|\int dx dy~|F_p|^2F^\ast_sF_i\Big|,
\label{eq:etaDef}
\end{equation}
where $F_p$, $F_s$, and $F_i$ are the mode field profiles evaluated at $\omegab_p$, $\omegab_s$, 
and $\omegab_i$, respectively. Each mode field profile is assumed to be normalized to unity, i.e. $\int dxdy~|F_j|^2=1$. 
$\eta_{ppsi}$ is the overlap integral of the different modes involved in the FWM process and determines the efficiency of 
a specific FWM process.

\begin{figure}[h]
\includegraphics[width=3.0 in]{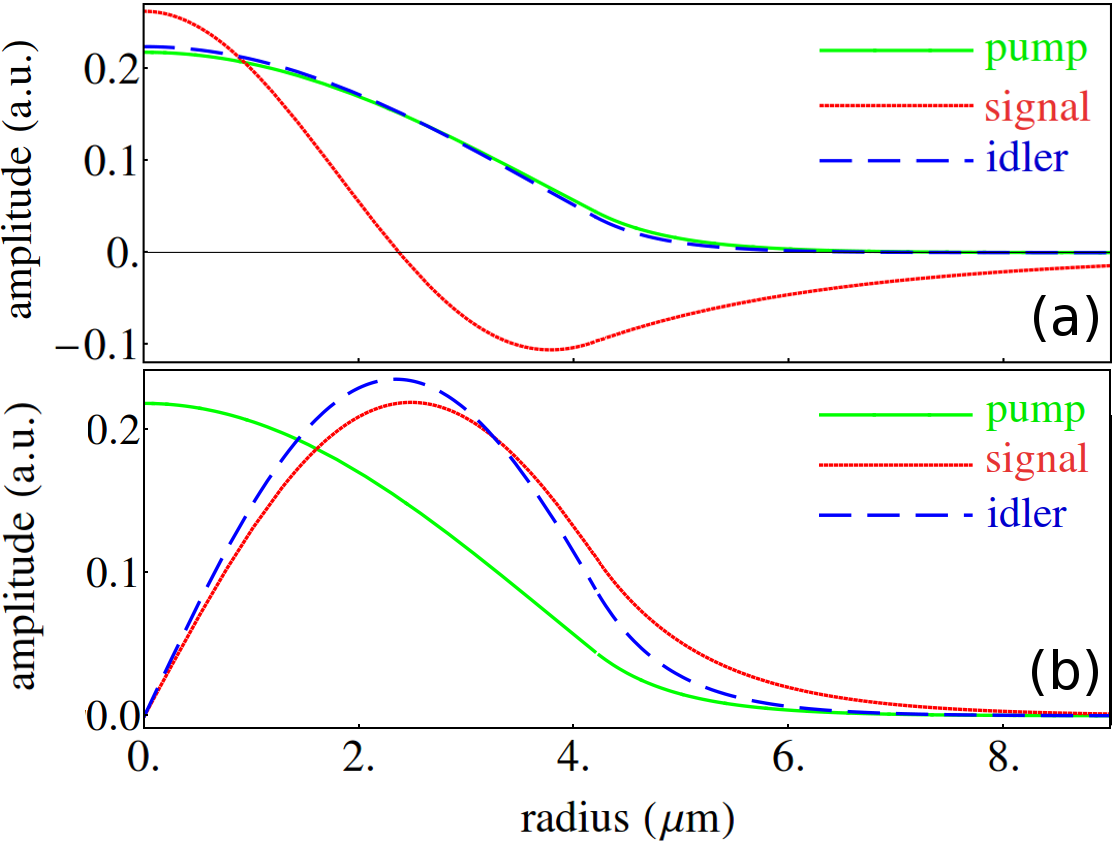}
\caption{Normalized radial mode profiles for processes $\mathcal{A}$ and $\mathcal{B}$ are shown in subfigures (a) and (b), respectively.}
\label{Fig:modes}
\end{figure}

In Fig.~\ref{Fig:modes}(a) and (b), we show the radial profiles of the modes in the two separate FWM processes 
$\mathcal{A}$ and $\mathcal{B}$, which were discussed in detail in subsection~\ref{sec:intermodal1}, respectively.   
The profiles are determined at their corresponding wavelengths, and clearly capture the reason behind different
efficiencies for different FWM processes determined by the overlap integral in Eq.~\ref{eq:etaDef}. 

\begin{figure}[h]
\includegraphics[width=3.4 in]{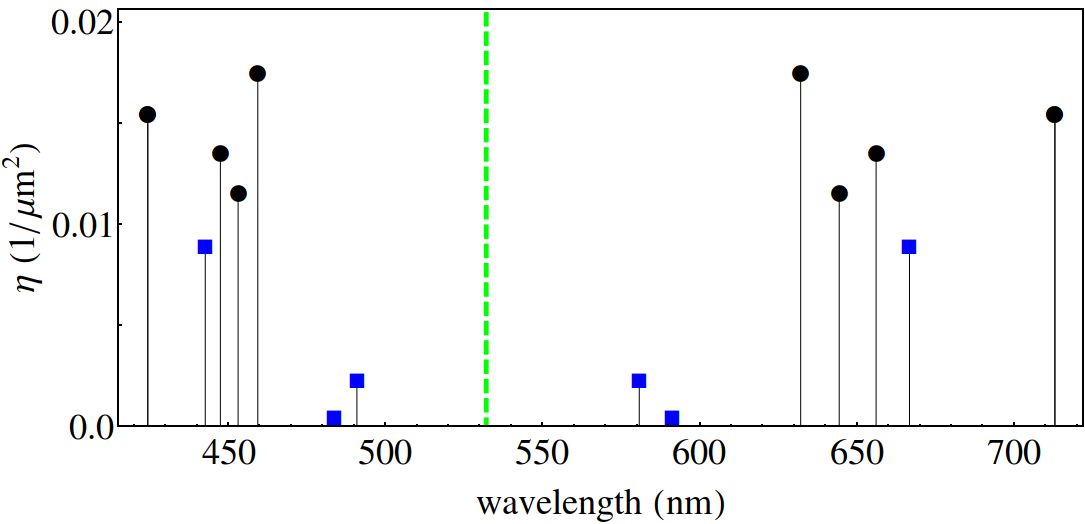}
\caption{Calculated FWM efficiencies for single-spatial-mode pump configurations. The black circles (blue squares) belong to the FWM processes for the pump photons 
in the ${\rm LP}_{01}$ (${\rm LP}_{11}$) mode. The FWM processes of $\mathcal{A}$ and $\mathcal{B}$ are observed in our experiments.}
\label{Fig:eta}
\end{figure}

In Table.~\ref{table:eta}, we have calculated $\eta$ for the phase-matched FWM processes which result in large spectral shifts 
relative to the SRS peak. The calculations are done by considering several different scenarios where the pump photons can reside in 
${\rm LP}_{01}$ and ${\rm LP}_{11}$ modes and the signal and idler photons can be generated in ${\rm LP}_{01}$, ${\rm LP}_{11}$, and ${\rm LP}_{02}$ modes. 
The signal and idler wavelengths, as well
as the idler-pump and pump-signal spectral separation $\Omega_{pm}$ and the absolute value of the mode overlap integral $|\eta_{ppsi}|$
are listed in this table. In order to easily compare the strengths, we also show the value of $|\eta_{ppsi}|$ in Fig.~\ref{Fig:eta}, 
indicating the FWM processes from Table.~\ref{table:eta}. The processes with the pump in the fundamental ${\rm LP}_{01}$ mode of the fiber 
are more efficient than the ones generated from pump photons in ${\rm LP}_{11}$ modes as can be seen in Table.~\ref{table:eta} and Fig.~\ref{Fig:eta}.
The more efficient FWM processes with ${\rm LP}_{01}$ pump mode are also favored by the conventional coupling of the pump laser to the
optical fiber using an objective lens. Therefore, the comparative results in Table.~\ref{table:eta} and Fig.~\ref{Fig:eta} can also
be seen in experiment where the $\mathcal{A}$ and $\mathcal{B}$ processes are far more efficiently generated in our experiments.
\begin{table}[ht]
\caption{The major FWM processes with large spectral shifts in SMF-28 with 532~nm wavelength pump are shown. The overlap integral 
$\eta_{ppsi}$ is calculated for these processes.} 
\centering 
\begin{tabular}{p{0.8 cm} p{1.0 cm} p{1.0 cm} p{0.7 cm} p{1.0 cm} p{0.7 cm} p{1.0 cm} p{0.9 cm}} 
\hline\hline 
Proc.         & Pump            & Signal         & $\lambda_s$   & Idler & $\lambda_i$ & $\Omega_{pm}$ & $|\eta_{ppsi}|$\\ [0.8ex] 
\hline 
$\mathcal{A}$ & ${\rm LP}_{01}$ & ${\rm LP}_{02}$& 656 & ${\rm LP}_{01}$& 447 & 3553 & 0.0136\\ 
$\mathcal{B}$ & ${\rm LP}_{01}$ & ${\rm LP}_{11}$& 632 & ${\rm LP}_{11}$& 459 & 2975 & 0.0175\\
$\mathcal{C}$ & ${\rm LP}_{01}$ & ${\rm LP}_{02}$& 712 & ${\rm LP}_{02}$& 424 & 4770 & 0.0155\\
$\mathcal{D}$ & ${\rm LP}_{01}$ & ${\rm LP}_{01}$& 644 & ${\rm LP}_{02}$& 453 & 3276 & 0.0116\\
$\mathcal{E}$ & ${\rm LP}_{11}$ & ${\rm LP}_{02}$& 667 & ${\rm LP}_{02}$& 442 & 3796 & 0.0090\\
$\mathcal{F}$ & ${\rm LP}_{11}$ & ${\rm LP}_{01}$& 580 & ${\rm LP}_{02}$& 490 & 1575 & 0.0024\\
$\mathcal{G}$ & ${\rm LP}_{11}$ & ${\rm LP}_{02}$& 591 & ${\rm LP}_{01}$& 483 & 1880 & 0.0005\\ [0.6ex] 
\hline 
\multicolumn{8}{l}{$\Omega_{pm}$ and $\eta_{ppsi}$ have units of 1/cm and 1/$\mu {\rm m}^2$, respectively.}\\
\multicolumn{8}{l}{The signal and idler wavelengths $\lambda_s$ and $\lambda_i$ are in {\rm (nm)}.}\\
\hline
\end{tabular}
\label{table:eta} 
\end{table}

We would like to emphasize that the efficiency of a FWM process in a multimode optical fiber is proportional to the value of $\eta$
and the wavelength-dependent third order nonlinear susceptibility $\chi^{(3)}$; and is inversely proportional to the spectral separation 
of the signal and idler from the pump ${\rm \Omega}_{pm}$, and the normalized average of the group velocity dispersion evaluated in the signal and idler
spatial modes at the pump wavelength ${\bar\beta}^{(2)}$ as defined in Ref.~\cite{2015-Mafi-Pourbeyram-OPEX}.
As an example, $\eta$ for process $\mathcal{C}$ is slightly larger than process $\mathcal{A}$. However, $\bar{\beta}^{(2)}\approx 2.3\times 10^{-5}$cm 
for both processes, while the ${\rm \Omega}_{pm}$ is larger for process $\mathcal{C}$. Therefore, $(\bar{\beta}^{(2)}{\rm \Omega}_{pm})^{-1}$ equals to 9.1 
and 12.2 for process $\mathcal{C}$ and $\mathcal{A}$, respectively. This means that although process $\mathcal{A}$ has a lower value of $\eta$ in 
Fig.~\ref{Fig:eta}, the other relevant factors make it slightly more efficient, as is also observed clearly in the experiment.
\section{multiple factorable photon-pairs}
\begin{figure}[h]
\includegraphics[width=3.3 in]{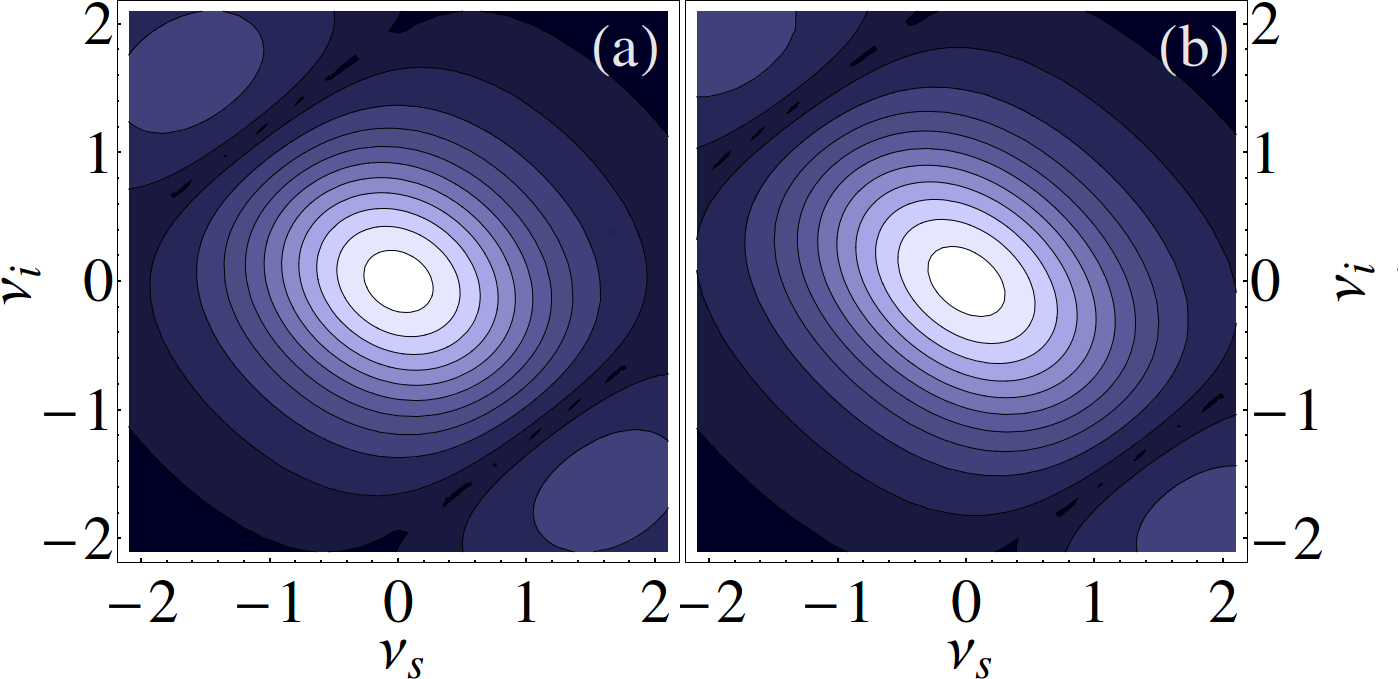}
\caption{Joint spectral amplitude for processes $\mathcal{A}$ and $\mathcal{B}$ are shown in (a) and (b), respectively.
The SMF-28 is 6~cm long and the pump wavelength is 532~nm with 1~THz bandwidth (equal to $1$~nm at 532.0 nm). 
The purity--calculated without the Gaussian approximation--for processes $\mathcal{A}$ and $\mathcal{B}$ is $84\%$ and $83\%$, respectively.
The purity will be 98\% and 94\%, respectively, if the sinc side-lobes are filtered out.}
\label{Fig:JSA}
\end{figure}

It was discussed earlier that one could generate a pure state by choosing a proper fiber length and pump bandwidth. We have found that by assuring the 
factorable state condition for one FWM process (e.g. process $\mathcal{A}$), the other one (process $\mathcal{B}$) would also become nearly factorable. 
Fig.~\ref{Fig:JSA} shows JSA for two FWM processes generated inside a 6~cm length of SMF-28 using a 532~nm pump with 1~nm bandwidth. The JSA for both
processes are nearly round, so both have high purity. The purity can also be increased by filtering the side-lobes of the sinc function~\cite{2013-Shapiro-Dixon}
 with minimal loss of flux as discussed before. This result shows that it is possible to generate multiple pure-state heralded photons in the same fiber using 
the same pump laser.
\section{widely tunable source of factorable two-photon states}
\begin{figure}[h]
\includegraphics[width=3.2 in]{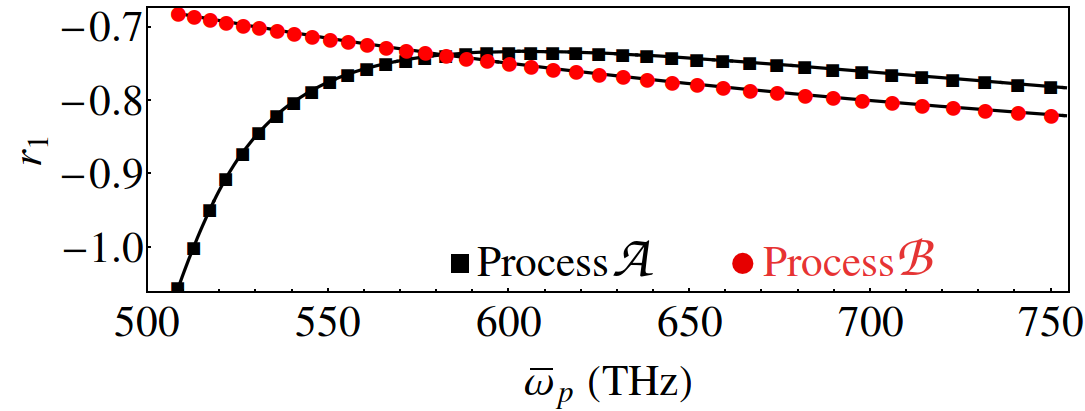}
\caption{$r_1$ as a function of the pump frequency from Eq.~\ref{Eq:r1}.}
\label{Fig:r1}
\end{figure}

\begin{figure}[h]
\includegraphics[width=3.2 in]{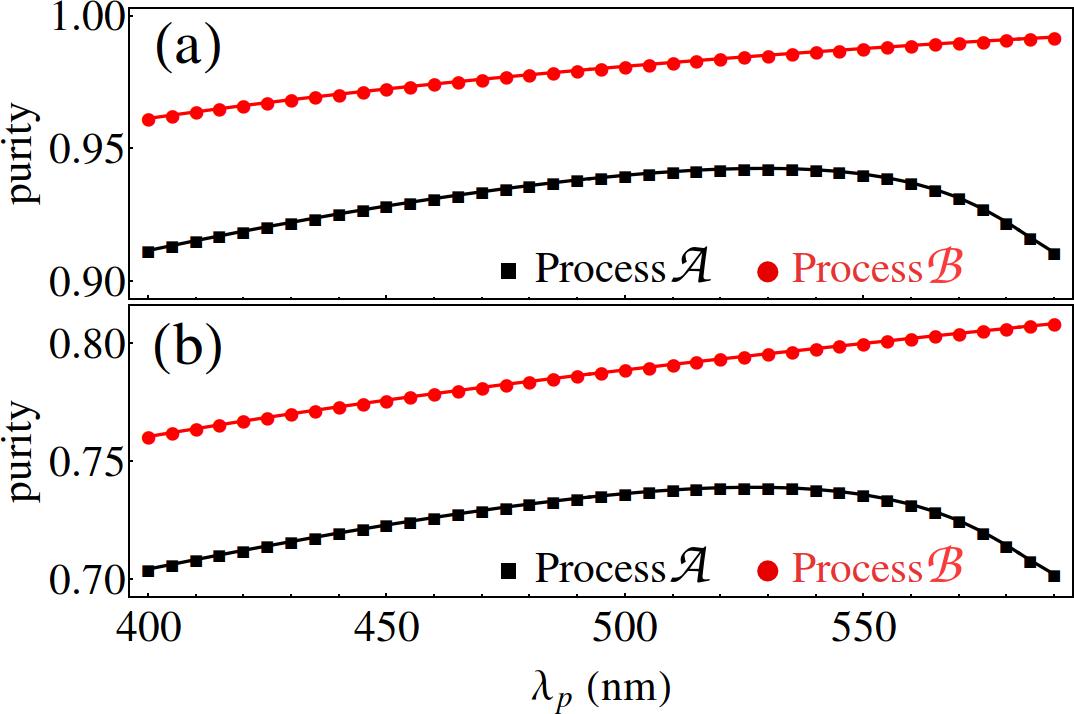}
\caption{Purity as a function of the pump wavelength (a) with and (b) without Gaussian approximation for two FWM processes
of $\mathcal{A}$ and $\mathcal{B}$ shown in black (square dots) and red (circular dots) colors, respectively. The fiber is 
10~cm long and and the pump bandwidth is 1~THz.}
\label{Fig:purity}
\end{figure}

Another interesting observation is that not only is it possible to simultaneously obtain high purity
photons from separate FWM processes as discussed in the previous section, the purity is also not 
affected much as the pump wavelength is varied over a wide range of more than 100~nm. 
This behavior can be understood from the phase-matching equation by taking a derivative 
of Eq.~\ref{Eq:phase-matching} with respect to $\omegab_p$ to obtain
\begin{equation}
\frac{2}{1-\tau_s/\tau_i}=\frac{d\omegab_s}{d\omegab_p}.
\label{Eq:r1}
\end{equation}
From Fig.~\ref{Fig:PM-smf28}, it can be seen that $d\omegab_s/d\omegab_p$ does not change appreciably over a wide pump wavelength range so
$d\omegab_s/d\omegab_p$ is nearly constant. Therefore, $\tau_s/\tau_i$ remains nearly constant over  a wide pump wavelength range. We recall
that $\tau_s/\tau_i$ is the $r_1$ parameter introduced in section~\ref{sec:purity} and is the horizontal axis in the purity contour plots
of Fig.~\ref{Fig:purity-r1-r2}. In Fig.~\ref{Fig:r1}, the change in the value of $r_1$ with the pump wavelength is plotted for processes $\mathcal{A}$ and $\mathcal{B}$.
Therefore, because changing the pump wavelength does not significantly affect the value of $r_1$ when the pump frequency
$\omegab_p$ is varied from 550~THz to 750~THz, the position of the system configuration on the purity contour plots 
in Fig.~\ref{Fig:purity-r1-r2} can only vary in the vertical direction. We have also shown that the variation in $r_2$ 
is small over this range, so the value of purity does not change appreciably when the pump wavelength is varied. This
can be clearly seen in Fig.~\ref{Fig:purity}, where Fig.~\ref{Fig:purity}(a) and (b) are with and without the Gaussian 
approximation of the sinc function, respectively. Moreover, because we argued that the purity values of 
processes $\mathcal{A}$ and $\mathcal{B}$ were nearly the same, they remain similar as the pump frequency is changed.  
\section{Discussion}
In this manuscript, we have presented a detailed account of the purity of a photon pair generated via intermodal 
phase matching in a multimode optical fiber and shown that it is possible to engineer nearly factorable photon-pair 
state. However, we emphasize that this is not the only way to generate factorable photon-pair states. For example,
it is possible to engineer the fiber dispersion or tune the pump bandwidth to generate factorable states in a  
single-mode optical fiber in which the photon pair is generated via near-ZDW phase-matched FWM process~\cite{2007-Walmsley-GarayPalmett,2009-Walmsley-Cohen}.
There are other ways to generate high purity photon-pair sates as well. For example, Ref.~\cite{2013-Lorenz-Fang}
uses two spectrally distinct pump pulses in optical fibers and shows that the group velocity difference 
between the pump pulses and their temporal walk-off results in a gradual turn-on and turn-off of the FWM interaction
that enables the creation of completely factorable states in properly designed systems. Birefringence in optical fibers 
has also been used to create factorable photon pairs~\cite{2009-Walmsley-Smith-1,2011-Walmsley-Soller}.

The main advantage of factorable photon-pair state generation via intermodal phase matching in a multimode optical 
fiber is the fact that the photon pairs are generated with large spectral separations from the pump wavelength, such 
that the photon pairs are immune to contamination from spontaneous Raman scattering and residual pump photons. In addition,
a multimode fiber can simultaneously support several phase-matched FWM process, resulting in multiple distinct 
photon-pair states. There are however several drawbacks to this method. For example, standard optical fiber 
communication systems rely on single-mode fibers supporting only ${\rm LP}_{01}$ modes. If the heralded photon is generated
in a different mode, it may bear a considerable coupling loss from the multimode to single-mode communication fiber. There
are ways to mitigate this problem, for example by using large period gratings to fully rotate the non-${\rm LP}_{01}$ mode
to ${\rm LP}_{01}$ in the multimode fiber~\cite{Sumetsky:08}. Another draw-back is the fact the photon-pair generation 
rate is inversely proportional to the spectral separation from the pump~\cite{2015-Mafi-Pourbeyram-OPEX}.

The discussion presented in this manuscript relies heavily on the integrity of modal propagation in the optical fiber. This 
however can be disturbed in the presence of unwanted mode coupling effects that are triggered by 
longitudinal inhomogeneities in the optical fiber. A possible source of variation can be the presence of scatterers in the 
optical fiber. Here we use Corning~SMF-28 which is a very high quality telecommunication-grade optical fiber and the coupling 
length among propagating modes because of impurities far exceeds the sub-meter lengths suggested in this manuscript. Another
source of linear mode coupling can be due to micro- and macro-bending effects. Again, the impact of micro-bending on mode-coupling
in sub-meter lengths is virtually zero. Marco-bending can result in coupling among the modes. To get a sense of the required bending
radius to cause significant mode coupling, we calculate the coupling coefficient $\mathcal{C}$ between ${\rm LP}_{01}$ and ${\rm LP}_{11}$ modes
when the fiber is bent at a radius of $R_c$ using the methods presented in Refs.~\cite{tagkey1991iii,Shemirani:09,2013-Mafi-Rodrigo} and show that
$\mathcal{C}=n_0\pi w/\sqrt{2}R_c\lambda$, where we use the same definition of $\mathcal{C}$ as in Ref.~\cite{saleh2007fundamentals}.
$w$ is the nominal mode-field radius of the ${\rm LP}_{01}$ and ${\rm LP}_{11}$ modes and $n_0$ is the core refractive index.
Efficient coupling is only possible if the coupling coefficient $\mathcal{C}$ is larger or at least comparable to the difference
between the propagation constants $\Delta\beta$ of ${\rm LP}_{01}$ and ${\rm LP}_{11}$ modes. It can be shown that 
$\Delta\beta\approx \lambda/2n_0a^2$, where $a$ is the fiber core radius. The condition for 50\%-level coupling between these modes
is $\mathcal{C}\approx \Delta\beta/2$ (see Ref.~\cite{saleh2007fundamentals}), which results in
\begin{align}
R_c\approx \dfrac{2\sqrt{2}n^2_0\pi w a^2}{\lambda^2}\approx 3.4{\rm mm}.
\end{align}
Therefore, the experimenters must be careful not to create tight bends but bending the fiber to an average radius of larger than a few 
centimeters will not cause any noticeable mode coupling. 

Finally, there exists the possibility for Kerr nonlinearity-induced mode coupling if the pump laser simultaneously
excites more than one mode in the fiber where the nonlinear change in the refractive index can potentially create an effective long-period grating
that affects the integrity of signal and idler modes through nonlinear cross-phase modulation. This effect is negligible for single-photon
generation because the pump power must be kept rather low to prevent stimulated photon generation; however, we will show in a subsequent 
paper that this effect can fully rotate modes in fiber lengths even shorter than one meter in experiments that use high pump power for
new wavelength generation using FWM in multimode fibers such as in the work presented in Ref.~\cite{2015-Mafi-Pourbeyram-OPEX}. 
\section{Conclusion}
We have studied theoretically the spontaneous four-wave mixing in a multimode optical fiber as a source of state engineered photon pairs. 
We have shown that it is possible to design this source so as to yield photon pairs with a broad class of spectrally engineered properties, 
including factorable and wavelength-tunable states. A major drawback of conventional fiber-based photon pair sources is contamination 
by SRS photons; this issue is fully addressed in the configuration presented here because the idler and signal photons can be generated
with spectral shifts that are much larger than the SRS bandwidth in silica fibers. This alleviates the need for extra filtering 
or cryogenic cooling of the fiber to suppress the SRS photons.
We have demonstrated the capability to tailor two-photon states through FWM and generate pure-state single-photon wavepackets in different 
fiber modes and through multiple simultaneous FWM processes. We have also shown the possibility of generating two separate FWM processes 
and state-engineering both to obtain simultaneous factorable states from both. We expect our results to be useful in designing sources for 
practical implementations of quantum information processing technologies.
\section{acknowledgement}
Hamed Pourbeyram and Arash Mafi acknowledge support by Grant Number 1522933 from the National Science Foundation.
\bibliography{blist}
\end{document}